\documentclass[preprint,showpacs,showkeys,amsmath,amssymb,aps,doublespace]{revtex4}
\usepackage[dvips]{graphicx}
\usepackage{setspace}
\usepackage{amsmath}
\usepackage{amsfonts}
\usepackage{amssymb,color}
\usepackage{graphicx,color}
\newcommand{\np}{\nu_\perp}
\begin{document}

\title{A symbiotic two-species contact process}

\author{Marcelo Martins de Oliveira$^1$\footnote{email: mmdeoliveira@ufsj.edu.br},
Renato Vieira Dos Santos$^2$,
and Ronald Dickman$^2$\footnote{email: dickman@fisica.ufmg.br}
}
\address{
$^1$Departamento de F\'{\i}sica e Matem\'atica, CAP, Universidade Federal de S\~ao Jo\~ao del Rei,
36420-000 Ouro Branco, Minas Gerais - Brazil\\
$^2$Departamento de F\'{\i}sica and National Institute of Science and Technology for Complex Systems,
ICEx, Universidade Federal de Minas Gerais, C. P. 702, 30123-970 Belo Horizonte, Minas Gerais - Brazil
}

\date{\today}

\begin{abstract}
We study a contact process (CP) with two species that interact in a symbiotic manner.
In our model, each site of a lattice may be vacant or host individuals of
species A and/or B; multiple occupancy by the same species is prohibited.
Symbiosis is represented by a reduced death rate, $\mu < 1$, for individuals
at sites with both species present.  Otherwise, the dynamics
is that of the basic CP, with creation (at vacant neighbor sites) at rate $\lambda$
and death of (isolated) individuals at a rate of unity.
Mean-field theory and Monte Carlo simulation show that the critical creation rate,
$\lambda_c (\mu)$, is a decreasing function of $\mu$, even though a single-species
population must go extinct for $\lambda < \lambda_c(1)$, the critical point
of the basic CP.
Extensive simulations yield results for critical behavior that are compatible
with the directed percolation (DP) universality class, but with unusually strong
corrections to scaling.  A field-theoretical argument
supports the conclusion of DP critical behavior.
We obtain similar results for
a CP with creation at second-neighbor
sites and enhanced survival at first neighbors, in the form of an
annihilation rate that decreases with the number
of occupied first neighbors.
\end{abstract}

\pacs{05.50.+q,05.70.Ln,05.70.Jk,02.50.Ey,87.23.Cc}

\maketitle

\section{Introduction}

Absorbing-state phase transitions have attracted
much interest in recent decades, as they appear in a wide
variety of problems, such as population dynamics,
heterogeneous catalysis, interface growth, and epidemic
spreading \cite{marro,munoz,henkel,odor07,odor04}. Interest in such
transitions has been further stimulated by recent experimental realizations \cite{take07,pine}.

The absorbing-state universality class associated with directed percolation (DP) has proven
to be particularly robust.  DP-like behavior appears to be generic for absorbing-state
transitions in models with short-range interactions and lacking a conserved density
or symmetry beyond translational invariance \cite{janssen,grassberger}.
By contrast, models possessing two
absorbing states linked by particle-hole symmetry belong to the voter model universality class \cite{voter}.

The contact process (CP) \cite{harris-CP} is probably the best understood model
exhibiting an absorbing-state phase transition; it has been known for many years
to belong to the DP class.  The CP can be interpreted as
a stochastic birth-and-death process with a spatial structure.  As a control
parameter (the reproduction rate $\lambda$) is varied, the system undergoes a phase
transition between extinction and survival.  In this context it is natural to
seek a manner to include symbiotic interactions in the CP.  In the present work,
this is done by allowing two CPs (designated as species A and B) to inhabit the
same lattice.  The two species interact via a reduced death rate, $\mu$,
at sites occupied by individuals of both species.  (Aside from this interaction,
the two populations evolve independently.)  We find, using mean-field theory
and Monte Carlo simulation, that the symbiotic interaction favors survival
of a mixed population, in that the critical reproduction rate $\lambda_c$ decreases
as we reduce $\mu$.  Note that for $\lambda(\mu) < \lambda < \lambda(1)$, {\it only}
mixed populations survive; in isolation, either species must go extinct.

In addition to its interest as a simple model of symbiosis, the critical
behavior of the two-species CP is intriguing in the context of nonequilibrium
universality classes.  By analogy with the (equilibrium) $n$-vector model,
in which the critical exponents depend on the number of spin components $n$,
one might imagine that the presence of two species
would modify the critical behavior.  Using extensive simulations,
we find that the critical behavior is consistent with that of directed percolation (DP),
although with surprisingly strong corrections to scaling.  An argument based on field theory
supports the conclusion of DP scaling.
We note that our result agrees with that of Janssen, who studied
general multi-species DP processes \cite{janssen01}.
Similar conclusions apply to a related model,
a CP with creation at second-neighbor
sites and enhanced survival at first neighbors, in the form of an
annihilation rate that decreases with the number
of occupied first neighbors.  (In this case the two species inhabit
distinct sublattices.)

The balance of this paper is organized as follows. In the next
section we define the models and analyze them using mean-field theory.
In Sec. III we present our simulation results, and in Sec. IV we discuss
a field-theoretic approach.  Sec. V is devoted to discussion and conclusions.

\section{Models and Mean-Field Theory}

To begin we review the definition of the basic contact process.
Following the usual nomenclature, we refer to an active site as being
occupied by a ``particle" and an inactive one as ``vacant".
The CP \cite{harris-CP} is a stochastic interacting particle
system defined on a lattice, with each site $i$ either occupied by a particle
[$\sigma_i (t)= 1$], or vacant [$\sigma_i (t)= 0$]. Transitions from
$\sigma_i = 1$ to $\sigma_i = 0$ occur at a rate of unity,
independent of the neighboring sites. The reverse transition, a vacant site becoming
occupied, is only
possible if at least one of its nearest neighbors (NNs) is occupied: the
transition from $\sigma_i = 0$ to $\sigma_i = 1$ occurs at rate
$\lambda r$, where $r$ is the fraction of NNs of site
$i$ that are occupied.  Thus the state $\sigma_i = 0$ for all $i$ is
absorbing.  At a certain critical value
$\lambda_c$ the system undergoes a phase transition
between the active and the absorbing state \cite{harris-CP}.  The CP has
been studied intensively via series expansion and
Monte Carlo simulation, and its critical properties are known to
high precision \cite{marro,henkel,odor07,hinrichsen,odor04}.

We now define a two-species symbiotic contact process (CP2S).  Let the indicator
variables for occupation of site $i$ by species A and B be
$\sigma_i$ and $\eta_i$, respectively.  The allowed states for a site, $(\sigma_i, \eta_i)$,
are $(0,0)$, $(0,1)$, $(1,0)$, and $(1,1)$.  The transitions
$(0,0) \to (0,1)$ and $(1,0) \to (1,1)$ occur at rate $\lambda r_A$ with $r_A$
the fraction of NNs bearing a particle of species A.  Similarly, the rate for
the transitions $(0,0) \to (1,0)$ and $(0,1) \to (1,1)$ is $\lambda r_B$
with $r_B$ the fraction of NNs bearing a particle of species B.
The transitions $(0,1) \to (0,0)$ and $(1,0) \to (0,0)$ occur at a rate of
unity, whereas $(1,1) \to (1,0)$ and $(1,1) \to (0,1)$ at rate $\mu$.
This set of transition rates describes a pair of contact processes inhabiting the
same lattice.  For $\mu=1$ the two processes
evolve independently, but for $\mu < 1$ they interact symbiotically, since
the annihilation rates are reduced at sites with both species present.
We note that the rates are symmetric under exchange of species labels A and B.

We also study a CP with creation at second-neighbor sites.
In Ref. \cite{cpsl} a modified CP was defined so:

i) In addition to creation at NNs, at rate $\lambda_1$, we allow
creation at {\it second neighbors}, at rate $\lambda_2$.  For bi-partite lattices such as the
ring or the
square lattice, $\lambda_1$ is the rate of creation in the
opposite sublattice, while $\lambda_2$ is the rate in the
same sublattice as the replicating particle.

ii) The annihilation rate at a given site is $1 + \nu n^2$, with $n$ denoting
the number of occupied NNs.

For $\nu > 0$, the presence of particle in one sublattice tends to suppress their
survival in the other, leading to the possibility of sublattice ordering, as
discussed in \cite{cpsl}.

Suppose now that $\lambda_1 = 0$, and let $\lambda_2 \equiv \lambda$.  Then the
populations in the two sublattices constitute distinct species, since creation
is always in the same sublattice.  For $\nu < 0$, moreover, the two species
interact in a symbiotic manner, analogous to that in the two-species CP
defined above.  (For $\nu = 0$ the two sublattices evolve independently.)
We call this process the symbiotic sublattice contact process (CPSLS).

Both the CP2S and CPSLS possess four phases: the fully active phase (nonzero populations
of both species), a symmetric pair of partly active phases (only one species present),
and the inactive phase (all sites inactive).  The latter is absorbing while the
partly active phases represent absorbing subspaces of the dynamics.  (That is, a species
cannot reappear once it goes extinct.)  Let $\lambda_{c,0}$ denote the critical
creation rate of the basic CP.  In the CP2S with $\mu=1$ (or the CPSLS with $\nu = 0$),
the critical creation rate must be $\lambda_{c,0}$.  The same applies for the
transitions from the partly active phases to the absorbing one, regardless
of the value of $\mu$ or $\nu$. Intuitively, in the
presence of symbiotic interactions, one expects the transition from the fully active
to the absorbing phase to occur at some $\lambda_c < \lambda_{c,0}$, since the annihilation
rate is reduced.  Since this expectation is borne out numerically, the partly active phases
are of little interest, as they are not viable in the vicinity of the fully active-absorbing
phase transition.  Understanding the latter transition is the principal objective of this study.

As a first step in characterizing the phase diagrams of the models, we develop mean-field
approaches.  The derivation of a dynamic mean-field theory (MFT) for an interacting
particle system begins with the master equation for the set of one-site probabilities (or,
more generally, the $n$-site joint probability distribution) \cite{marro}.
In this equation, the $n$-site probability
distribution is inevitably coupled to the distribution for $n+1$ or more sites.  An $n$-site MFT is
obtained by estimating the latter distribution(s) in terms of that for $n$ sites.  Here we
consider the simplest cases, $n=1$ and 2.

Consider the CP2S in the one-site approximation.  Denoting the probabilities for a given site to be
vacant, occupied by species A only, by species B only, and doubly occupied by $p_0$, $p_A$, $p_B$,
and $P_{AB}$, respectively, assuming spatial homogeneity, and factorizing two-site joint
probabilities ($p[(\sigma_i,\eta_i),(\sigma_j,\eta_j)] = p[(\sigma_i,\eta_i)]p[(\sigma_j,\eta_j)]$)
one readily obtains the equations

\begin{eqnarray}
\frac{d p_0}{dt} &=& - \lambda p_0 (\rho_A + \rho_B) + p_A + p_B,
\nonumber
\\
\frac{d p_A}{dt} &=&  \lambda p_0 \rho_A + \mu p_{AB} - (1 + \lambda \rho_B) p_A,
\nonumber
\\
\frac{d p_B}{dt} &=&  \lambda p_0 \rho_B + \mu p_{AB} - (1 + \lambda \rho_A) p_B,
\nonumber
\\
\frac{d p_{AB}}{dt} &=& \lambda (p_A \rho_B + p_B \rho_A) - 2 \mu p_{AB},
\end{eqnarray}

\noindent where $\rho_A = p_A + p_{AB}$ and $\rho_B = p_B + p_{AB}$.  If one species
is absent (so that, say, $p_B = p_{AB} = 0$) this system reduces to the MFT
for the basic contact process, $\dot{p}_A = \lambda p_A (1-p_A) - p_A$, with a
critical point at $\lambda = 1$.  To study the effect of symbiosis we seek a
symmetric solution, $p_A = p_B = p$.  In this case one readily finds
the stationary solution:

\begin{equation}
\overline{p} = \frac{\mu}{2 \lambda (1-\mu)} \left[ 2(1-\mu) - \lambda
+ \sqrt{\lambda^2 - 4\mu (1-\mu)} \right].
\label{pMFT}
\end{equation}
and
\begin{equation}
\overline{p}_{AB} = \frac{\lambda p^2}{\mu - \lambda p}
\label{pabmft}
\end{equation}

\begin{figure}[!hbt]
\includegraphics[clip,angle=0,width=0.8\hsize]{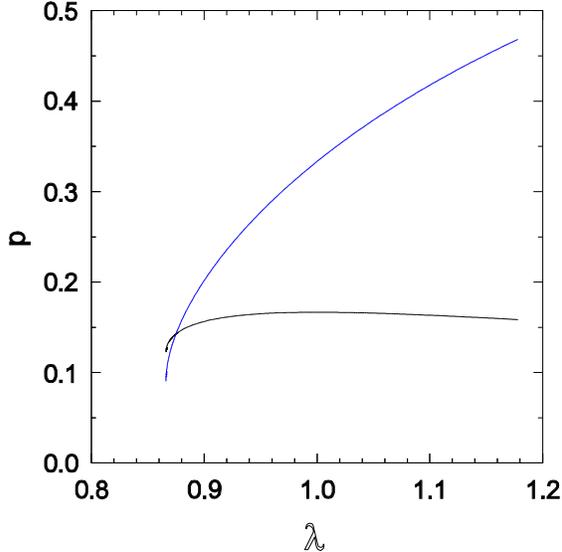}
\caption{\footnotesize{Density $p$ of species A (lower curve) and of doubly occupied
sites, $p_{AB}$ (upper curve) in the one-site approximation for the CP2S, $\mu=0.25$.
}}
\label{c25}
\end{figure}

\noindent For $\mu \geq 1/2$, $p$ grows continuously from
zero at $\lambda=1$, marking the latter value as the critical point.
The activity grows linearly, $p \simeq [\mu/(2\mu -1)](\lambda-1)$, in this regime.
For $\mu < 1/2$, however, the expression is already positive for
$\lambda = \sqrt{4 \mu(1-\mu)} < 1$, and there is a {\it discontinuous}
transition at this point.  The value $\mu = 1/2$ may be viewed as a
tricritical point; here $p \sim \sqrt{\lambda-1}$ for $\lambda > 1$.
Numerical integration of the MFT equations confirms the above results.
For $\mu < 1/2$, MFT in fact furnishes the {\it spinodal} values of $\lambda$.
For a given set of initial probabilities, the numerical integration converges to
the active stationary solution for $\lambda \geq \lambda^*$ and to the
absorbing state for smaller values of $\lambda$.  For the most favorable
initial condition, i.e., $p_{AB}(0) \to 1$, $\lambda^* \to \lambda^{(-)} = \sqrt{4 \mu(1-\mu)}$, the
lower spinodal, while for a vanishing initial activity, $\rho_A$, $\rho_B \to 0$,
$\lambda^* \to \lambda^{(+)} = 1$.  The stationary activity at $\lambda^*$ is
nonzero.  Fig. \ref{c25} shows the stationary probabilities versus $\lambda$ for
$\mu = 1/4$.

\begin{figure}[!hbt]
\includegraphics[clip,angle=0,width=0.8\hsize]{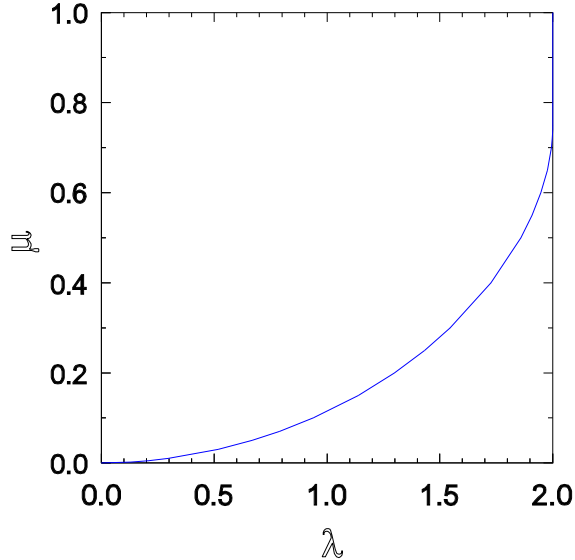}
\caption{\footnotesize{Phase boundary in the $\lambda-\mu$ plane as given by two-site MFT
for the CP2S on the line.  The curved portion represents the lower spinodal, $\lambda^{(-)} (\mu)$.
}}
\label{cp2spa}
\end{figure}

The two-site MFT for the one-dimensional CP2S involves ten pair probabilities and a set of thirty-two
transitions.  The resulting phase diagram is qualitatively similar to that of the
one-site MFT.  For $\mu > 0.75$, the transition is continuous and occurs at $\lambda = 2$,
the same value as for the basic CP at this level of approximation.  There is a tricritical
point at $\mu = 0.75$, below which the transition is discontinuous; Fig. \ref{cp2spa}
shows the phase phase boundary.

The one-site mean-field theory (MFT) for the CPSL was developed in Ref. \cite{cpsl}.  Adapted
to the present case (creation only in the same sublattice, symbiotic interaction),
the equation is

\begin{equation}
\frac{d \rho_A}{dt} = - (1 - \nu q^2 \rho_B^2) \rho_A + \lambda_2 \rho_A (1-\rho_A)
\end{equation}

\noindent and similarly for $\rho_A \leftrightharpoons \rho_B$,
on a lattice of coordination number $q$.  (Here $\rho_j$ denotes the
fraction of occupied sites in sublattice $j$.)
As we seek a symmetric solution, we set $\rho_A = \rho_B$.  The resulting
equation yields a continuous phase transition at $\lambda = 1$, independent of
$\nu$.  (Note that $\nu$ must be greater than $-1/16$; smaller values correspond to a
{\it negative} annihilation rate, for $\rho$ near unity.)
The two-site approximation is likely to provide a better description of
the CPSLS, since in this case the nearest-neighbor double occupancy
probability is an independent variable, analogous to $p_{AB}$ in the one-site
MFT of the CP2S.  Since such an analysis is unlikely to result in additional
insights, we shall not pursue it here.

Although MFT predicts
a discontinuous phase transition in the CP2S in any number of dimensions,
such a transition is not possible in one-dimensional
systems with short-range interactions and free of boundary fields \cite{hinrichsen1d}.
In one dimension the active-absorbing transition should be continuous, as we have
indeed verified in simulations.  Although our simulations show no evidence of a discontinuous
transition in two dimensions ($d=2$), such a transition remains a possibility for
$d \geq 2$, for small values of $\mu$.  A discontinuous transition might also arise
under rapid particle diffusion, as this generally favors mean-field-like behavior.

\section{Simulations}

We performed extensive Monte Carlo simulations of the CP2S on rings
and on the square lattice (with periodic boundaries), and of the CPSLS on rings.
A general observation is that both models appear to be more
strongly affected by finite-size corrections than is the basic CP.

In the simulation algorithm for the two-species CP, we maintain two lists, i.e., of singly
and doubly occupied sites.  Let $N_s$ and $N_d$ denote, respectively, the numbers of such
sites, so that $N_p = N_s + 2 N_d$ is the number of particles.  The total rate of (attempted)
transitions is $\lambda N_p + N_s + 2\mu N_d \equiv 1/\Delta t$, where $\Delta t$ is the
time increment associated with a given step in the simulation.  At each such step, we choose
among the events: (1) creation attempt by an isolated particle, with probability $\lambda N_s \Delta t$;
(2) creation attempt by a particle at a doubly occupied site, with probability $2 \lambda N_d \Delta t$;
(3) annihilation of an isolated particle, with probability $N_s \Delta t$; and (4)
annihilation of a particle at a doubly occupied site, with probability $2 \mu N_d$.
Once the event type is selected we choose a site $i$ from the appropriate list.    In case of annihilation, a
particle is simply removed, while creation requires the choice of a neighbor, $j$, of site $i$, and
can only proceed if $j$ is not already occupied by a particle of the species to be created.
For creation by a particle at a doubly occupied site, the species of the daughter particle
is chosen to be A or B with equal probability, and similarly for annihilation at a doubly
occupied site.

In simulations of the CPSLS we maintain a list of occupied
sites.  At each step a site is selected from the list; an attempt to create a new particle, at one of the
second-neighbor sites, is chosen with probability $p=\lambda/(1+\lambda_2+\mu n_1^2)$; the site
is vacated with the complementary probability, $1- p$.  The time increment associated with
each event is $\Delta t = 1/N_p$, with $N_p$ the number of particles just prior to the event.

\subsection{Results: CP2S in one dimension}

We studied the CP2S using three values of $\mu$: 0.9, 0.75, and 0.25.  While the first case may be seen as a
relatively small perturbation of the usual CP ($\mu=1$), the third represents a very
strong departure from the original model.  We perform three kinds of studies:
quasi-stationary (QS) \cite{qssim},
initial decay (starting from a maximally active configuration), and spreading, in which the
initial condition is a doubly occupied site in an otherwise empty lattice.  Although
the critical value, $\lambda_c (\mu)$, can be estimated using each method, spreading simulations
proved the most effective in this regard.

In the QS simulations, we study system sizes 800, 1600, 3200, 6400, and 12800, with each
run lasting 10$^7$ time units; averages and uncertainties are calculated over 10 - 80 runs.
We use three well established criteria to estimate the critical value:
(1) power-law dependence of the order parameter on system size,
$\rho \sim L^{-\beta/\nu_\perp}$,
and (2) of the lifetime, $\tau\sim L^{z}$, as well as (3) convergence of the
moment ratio, $m_\rho (L)$, to a finite limit, $m_c$, as $L \to \infty$ \cite{dic-jaf}.
Here $m_\rho \equiv \langle \rho^2 \rangle/\langle \rho \rangle ^2$.
The order parameter is defined as the density of individuals, i.e., $\rho = (N_A + N_B)/L$.
A related quantity of interest is the density $q$ of doubly occupied sites; the moment
ratio $m_q$ is defined in a manner analogous to $m_\rho$.
Two further quantities of interest are the scaled variances of $\rho$ and $q$;  we define
$\chi_\rho \equiv L^d \, \mbox{var}(\rho)$ and similarly for $\chi_q$.  The expected critical
behavior is $\chi \sim L^{\gamma/\nu_\perp}$, where the critical exponent $\gamma$ satisfies the
hyperscaling relation $\gamma = d \nu_\perp - 2 \beta$ \cite{marro}.

\begin{figure}[!hbt]
\includegraphics[clip,angle=0,width=0.8\hsize]{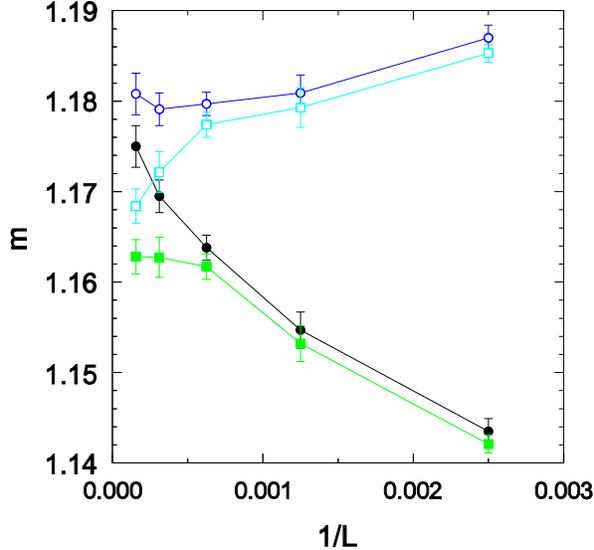}
\caption{\footnotesize{(Color online) QS Simulation of one-dimensional CP2S:
moment ratios $m_\rho$ (filled symbols) and $m_q$ (open symbols)
versus $1/L$ for the one-dimensional model with $\mu = 0.75$.  The upper curve in each pair is
for $\lambda = 3.0336$, the lower for $\lambda = 3.0337$.
}}
\label{m75}
\end{figure}

A preliminary estimate of $\lambda_c$ is obtained from the crossings of $m_\rho$ for successive
system sizes, $L$ and $2L$.  For $\mu=0.75$, for example, this yields $\lambda_c = 3.0337$.  The
plot of $m_\rho$ and $m_q$ (see Fig. \ref{m75}) indicates that $\lambda_c > 3.0336$ (since
$m_\rho$ curves upward), while the slight downward curvature for $\lambda = 3.0037$ suggests that
this value may be slightly above critical.  This graph also suggests that $m_\rho$ and $m_q$
approach the same limiting value, despite marked differences for smaller system sizes.
Table \ref{cp2s1qs} summarizes our findings for the critical parameters obtained from QS
simulations.

\begin{table}[h]
\centering \caption{CP2S in one dimension: results from
QS simulations, $L=800$, 1600, 3200, 6400, and 12800. For
$\mu=0.25$ the maximum size is 6400.
}
\label{cp2s1qs}
\begin{center}
\begin{tabular}{|c|c|c|c|c|c|c|c|} \hline
$\mu$  & $\lambda_{c}$ & $\beta/\np$ & $z$      &  $m_\rho$  & $m_q$ & ($\gamma/\np)_\rho$ & $(\gamma/\np)_q$  \\
\hline\hline
0.9    & 3.2273(1)     & 0.25(2)       & 1.50(5)   &   1.168(12) & 1.164(4) &  0.627(20) & 0.474(7) \\
0.75   & 3.03370(5)    & 0.241(6)      & 1.64(5)   &   1.163(10) & 1.166(2) &  0.528(6)  & 0.486(1) \\
0.25   & 1.76297(1)    & 0.248(3)      & 1.56(4)   &   1.168(3)  & 1.169(3) &  0.500(1)  & 0.492(2) \\ \hline\hline
CP/DP  & 3.29785       & 0.25208(5)    & 1.5807(1) &   1.1736(1) &          &  0.49584(9) &   \\ \hline
\end{tabular}
\end{center}
\end{table}
\vspace{2em}

\begin{figure}[!hbt]
\includegraphics[clip,angle=0,width=0.8\hsize]{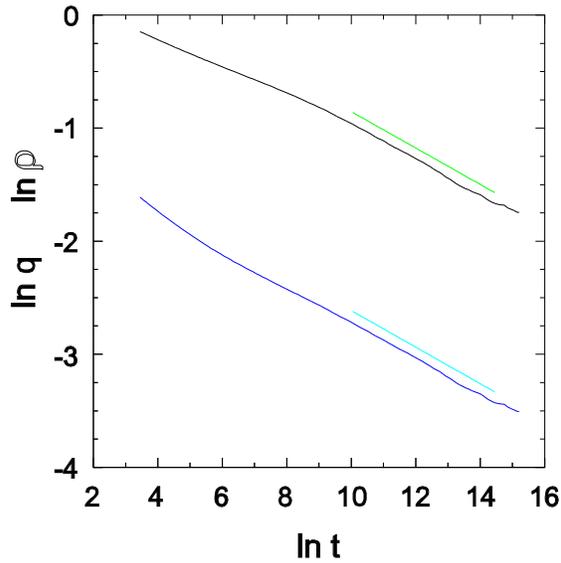}
\caption{\footnotesize{(Color online) Initial-decay simulation of CP2S in one dimension:
decay of the particle density $\rho$ (upper curve) and the density $q$
of doubly-occupied sites in initial decay studies with $\mu = 0.9$,
$\lambda = 3.2273$, system size $L=51200$.
The slopes of the regression lines are -0.161 ($\rho$) and -0.162 ($q$).
}}
\label{rho273}
\end{figure}

\begin{figure}[!hbt]
\includegraphics[clip,angle=0,width=0.8\hsize]{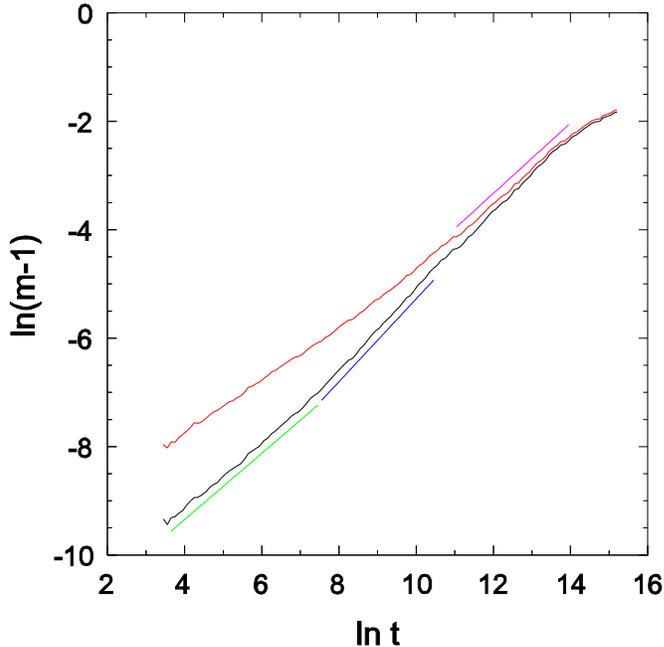}
\caption{\footnotesize{(Color online) Initial-decay simulation of CP2S in one dimension:
growth of fluctuations in $\rho$ (lower curve) and $q$
(upper curve) for parameters as in Fig. \ref{rho273}.
The slopes of the regression lines are (left to right): 0.613, 0.694, and 0.649.
}}
\label{zid273}
\end{figure}

The initial decay studies use, as noted above, an initial configuration with all sites doubly
occupied.  The activity then decays, following a power law, $\rho \sim t^{-\delta}$, at the
critical point \cite{torre}, until it saturates at its QS value.  The larger the system size,
the longer the period of power-law decay, and the more precise the resulting estimate for
the critical exponent $\delta$; here we use $L=25600$ or 51200.  Averages are calculated over
500-3000 realizations.  As the order parameter decays,
its fluctuations build up; at the critical point, the moment ratio is expected to
follow $m - 1 \sim t^{1/z}$ \cite{dasilva}.  Since we expect $\rho$ and $q$ to scale in the same
manner, we define exponents $\delta_\rho$ and $\delta_q$, and, similarly, $z_\rho$ and $z_q$,
based on the behavior of $m_\rho$ and $m_q$, respectively.  Figure \ref{rho273}, for
$\mu = 0.9$, shows that $\rho$ and $q$ decay in an analogous manner, and follow power laws at long
times, although there are significant
deviations from a simple power law at short times; the decay exponents are consistent with the
value of $\delta$ for directed percolation in one space and one time dimension (see Table \ref{cp2sid}).
The growth of fluctuations follows a more complicated pattern, as shown in Fig. \ref{zid273}.
At relatively short times, $m_\rho -1 \sim t^{1/z_\rho}$, with $z_\rho = 1.63(2)$, not very different
from the DP value; $m_q - 1 $ also grows as a power law in this regime, but with an
apparent exponent of $z_q = 2.06(1)$.  At longer times $z_\rho$ appears to take a smaller value
(1.31(1) for $7.5 < \ln t < 10.5)$, while $z_q$ shifts to a value close to that of DP (1.61(1) for
$10 < \ln t < 14$).  The reason for the distinct behaviors of $m_\rho$
and $m_q$, in marked contrast with the similar scaling of $\rho (t)$ and $q(t)$, is unclear.
While scaling anomalies are observed in the initial decay studies for $\mu=0.9$
and 0.75, for strong symbiosis ($\mu=0.25$) they are absent, as seen in Table \ref{cp2sid},
which summarizes the
results of the initial decay studies. (In this table, the values listed for $z_\rho$ and
$z_q$ reflect the latter part of the evolution,
during which the order parameter decays in the expected manner.)

\begin{table}[h]
\centering \caption{CP2S in one dimension: results from initial-decay studies.
}
\label{cp2sid}
\begin{center}
\begin{tabular}{|c|c|c|c|c|c|} \hline
$\mu$  & L     & $\lambda_{c}$  & $\delta$   & $z_\rho$    & $z_q$                   \\ \hline\hline
0.9    & 51200 & 3.2273         & 0.161(1)   &  1.44(1)    & 1.54(2)                 \\
0.75   & 25600 & 3.0337         & 0.1625(10) &  1.48(4)    & 1.55(4)                 \\
0.25   & 51200 & 1.76297        & 0.1581(3)  &  1.56(1)    & 1.58(1)                 \\
\hline\hline
 DP    &       &                & 0.1599     & 1.5807(1)   &                         \\ \hline
\end{tabular}
\end{center}
\end{table}
\vspace{2em}

In the spreading studies, each realization runs to a maximum time of $t_m$ (unless it falls into
the absorbing state prior to this).  The system size is taken large enough so that activity never
reaches the boundary.  Here we use $t_m = 2 \times 10^6$ and $L = 10^5$; averages are calculated over
$10^4$ or $2 \times 10^4$ realizations.
At the critical point, one expects
to observe power-law behavior of the survival probability, $P(t) \sim t^{-\delta}$, the mean
number of particles, $n(t) \sim t^\eta$, and the mean-square distance of particles from the
initial seed, $R^2 (t) \sim t^{z_s}$ \cite{torre}.  Here $\delta$ is the same exponent as governs the
initial decay of the activity, and $z_s$ is related to the dynamic exponent
$z$ via $z_s = 2/z$.  Deviations from asymptotic power laws, indicating
off-critical values of the control parameter $\lambda$, are readily identified in spreading simulations,
leading to precise estimates for $\lambda_c$.

\begin{figure}[!hbt]
\includegraphics[clip,angle=0,width=0.8\hsize]{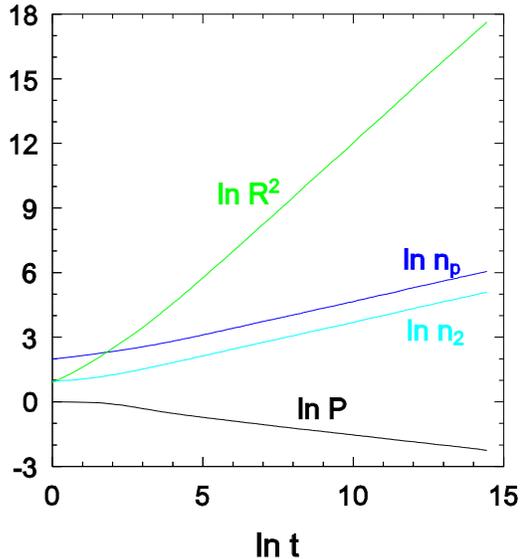}
\caption{\footnotesize{(Color online) Spreading simulation of CP2S in one dimension:
survival probability $P(t)$, total particle number
$n_p(t)$, number of doubly occupied sites, $n_2(t)$, and mean-square distance from seed, $R^2(t)$.
Parameters $\mu=0.25$, $\lambda = 1.76297$.
}}
\label{sp297}
\end{figure}

\begin{figure}[!hbt]
\includegraphics[clip,angle=0,width=0.8\hsize]{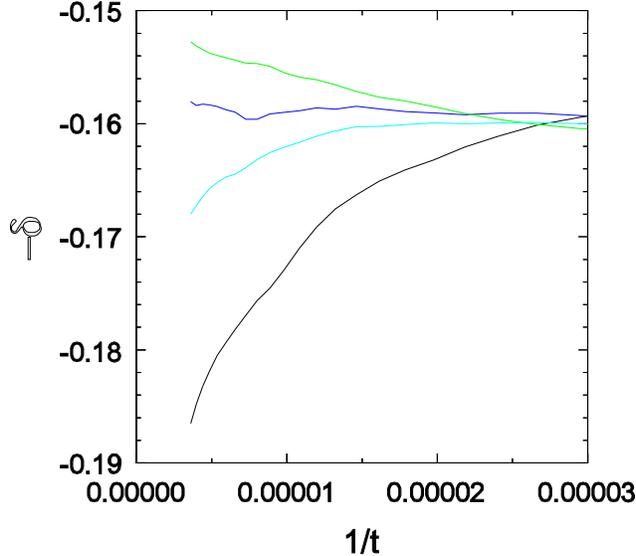}
\caption{\footnotesize{(Color online) Spreading simulation of CP2S in one dimension:
local slope $\delta(t)$ versus $1/t$ for
$\mu=0.25$ and (lower to upper) $\lambda = 1.7629$, 1.76295, 1.76297, and 1.7630.
}}
\label{del297}
\end{figure}

The spreading behavior is characterized by clean power laws, as illustrated in Fig. \ref{sp297}.
As this plot makes clear, the mean particle number, $n_p$, and the mean number of doubly occupied
sites, $n_2$, grow with the same critical exponent.  Precise estimates of the spreading exponents
are obtained via analysis of local slopes such as
$\delta(t)$, defined as the inclination of a least-square linear fit to the data (on logarithmic scales),
on the interval $[t/a, \,at]$.  (The choice of the factor $a$ represents a compromise between
high resolution, for smaller $a$, and insensitivity to fluctuations, for larger values;
here we use $a = 4.59$.)  Curvature in a plot of a local slope versus $1/t$ signals
an off-critical value.  Figure \ref{del297} shows the behavior of $\delta (t)$ for $\mu = 0.25$.
The spreading exponents, summarized in Table \ref{cp2spr}, are in good agreement with the
values for DP in 1+1 dimensions.  (We note that in all three cases, $\eta_p = \eta_2$ to within
uncertainty.)

\begin{table}[h]
\centering \caption{One-dimensional CP2S: Results from spreading simulations.
}
\label{cp2spr}
\begin{center}
\begin{tabular}{|c|c|c|c|c|} \hline
$\mu$ &    $\lambda_{c}$   & $\delta$  & $\eta$     & $z_{s}$    \\ \hline\hline
0.9   &       3.2273       & 0.165(1)  & 0.310(1)   & 1.257(2)    \\
0.75  &       3.0337       & 0.1595(5) & 0.3180(5)  & 1.265(5)    \\
0.25  &       1.76297      & 0.158(1)  & 0.315(3)   & 1.265(10)   \\
\hline\hline
 DP   &       3.29785      & 0.15947(5)& 0.31368(4) & 1.26523(3) \\ \hline
\end{tabular}
\end{center}
\end{table}
\vspace{2em}

\subsection{Contact process with creation at second-neighbors}

We studied the CPSLS using QS and initial decay simulations.  The results from the former, based on
FSS analysis of studies using $L=800$, 1600, 3200, 6400, and 12800, are summarized in Table \ref{cpsls1}.
The value of $\np$ was estimated (for $\nu = -0.1$ only) via analysis of the derivatives
$|dm/d\lambda|$, $d \ln \tau/d\lambda$ and $d \ln \rho_{p}/d\lambda$
in the neighborhood of the critical point.
Finite-size scaling implies that the derivatives
follow $| dx /dp| \propto L^{1/\nu_\perp}$ (here $x$ stands for any of the quantities mentioned).
We estimate the derivatives via least-squares linear fits to the data on an interval
that includes $\lambda_c$. (The intervals are small enough that the graphs show no significant
curvature.)  Linear fits to the data for $m$, $\ln \rho_p$,
and $\ln \tau$ yield
$1/\nu_\perp = 0.94(2)$, 0.92(3), and (again) 0.92(3), respectively, leading to the estimate
$\nu_\perp = 1.08(3)$.

Results of the initial decay studies are summarized in Table \ref{cpsls2}.  As in the
two-species CP, the value of $z$ obtained from analysis of $m(t)$ appears to be smaller than
the DP value, whereas the result obtained from QS simulations is consistent
with that of DP.

\begin{table}[h]
\centering \caption{One-dimensional CPSLS: results from quasistationary simulations.
}
\label{cpsls1}
\begin{center}
\begin{tabular}{|c|c|c|c|c|c|} \hline
$\nu$  & $\lambda_{c}$ & $\beta/\np$ & $z$       &    $m_c$     &      $\np$    \\ \hline\hline
-0.05  & 3.1489(1)     & 0.235(8)    & 1.63(5)   &   1.154(5)   &               \\
-0.1   & 2.8878(1)     & 0.242(1)    & 1.612(12) &   1.161(3)   &       1.08(3) \\
-0.2   & 2.0502(1)     & 0.253(6)    & 1.59(1)   &   1.170(6)   &               \\ \hline\hline
 DP    & 3.29785       & 0.25208(5)  & 1.5807(1) &   1.1736(1)  &   1.096854(4) \\ \hline
\end{tabular}
\end{center}
\end{table}
\vspace{2em}

\begin{table}[h]
\centering \caption{One-dimensional CPSLS: results from initial decay simulations.
}
\label{cpsls2}
\begin{center}
\begin{tabular}{|c|c|c|c|c|} \hline
$\mu$  & L     & $\lambda_{c}$ & $\delta$          & $z$    \\ \hline\hline
-0.05  & 50000 & 3.1489          & 0.1458(5)       &  1.45(2)       \\
-0.1   & 50000 & 2.8878          & 0.1484(7)       &  1.45(3)        \\
-0.2   & 20000 & 2.0503          & 0.1597(3)       &  1.53(1)        \\ \hline\hline
 DP    &       &                 & 0.1599          & 1.5807(1)                    \\ \hline
\end{tabular}
\end{center}
\end{table}
\vspace{2em}

\subsection{Two-species contact process in two dimensions}

\begin{figure}[h]
\includegraphics[width=8cm,clip=true]{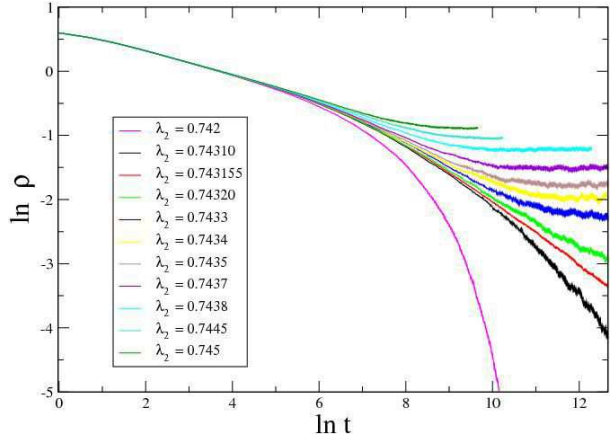}
\caption{\footnotesize{(Color online) CP2S in two dimensions:
Density of active sites starting from a maximally active initial condition,
for $\mu=0.1$, and $\lambda$ values as indicated. System size: $L=4000$.}}
\label{dmu01}
\end{figure}

\begin{figure}[h]
\includegraphics[width=8cm,clip=true]{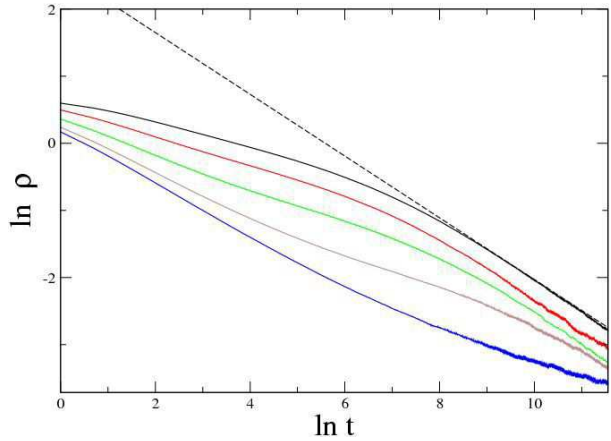}
\caption{\footnotesize{(Color online) CP2S in two dimensions:
density of active sites starting from a maximally active initial condition,
for $\mu=0.1, 0.25, 0.5, 0.75$ and $0.9$ (from top to bottom) and $\lambda = \lambda_c(\mu)$
(see Table \ref{cp2s2}). The slope of the dashed line is -0.45. System
size: $L=4000$.}}
\label{dmuall}
\end{figure}

We performed extensive Monte Carlo simulations of the CP2S on square lattices using both
initial decay and quasi-stationary (QS) simulations.
In order to locate the critical point with good precision, we study the initial decay of the
particle density, starting from a maximally active initial condition (all sites doubly occupied).
We use lattices
of linear size $L=4000$, and average over at least 20 different realizations.
Figure \ref{dmu01} shows the decay
of $\rho(t)$ for $\mu=0.1$. After an initial transient, during which the density evolves
slowly, the particle density follows a power law with $\delta=0.46(1)$, compatible with the
value ($\delta=0.4523(10)$) for the DP class in 2+1 dimensions. The transient behavior lasts
longer, the larger is $\mu$, as shown in Fig. \ref{dmuall}. However the relaxation is seen to cross over to
DP-like behavior for all values studied, except for $\mu=0.9$, for which the transient regime
persists throughout the entire simulation.

\begin{figure}[h]
\includegraphics[width=8cm,clip=true]{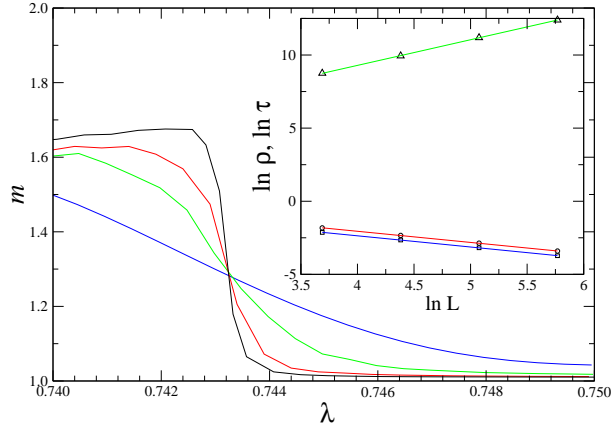}
\caption{\footnotesize{(Color online) CP2S in two dimensions: QS moment ratio of particles vs. $\lambda$. ,
for $\mu=0.1$ (System sizes: $L= 40, 80, 160, 320$
in order of steepness). Inset: QS density of active sites (circles),
density of doubly occupied sites (squares) and lifetime of the QS state (triangles), for $\mu=0.1$.}}
\label{scpQS}
\end{figure}

Having determined $\lambda_c$ to good precision in the initial decay studies,
we perform QS simulations of
the model on square lattices of linear size $L=20, 40,..., 320$ with periodic boundaries. Figure \ref{scpQS}
shows moment-ratio crossings and the finite-size scaling behavior of the density
and lifetime for $\mu=0.1$. For the larger sizes we
obtain $\beta/\nu_\perp= 0.78(1)$ and $z=1.74(2) $, in
good agreement with the best estimates for DP in 2+1 dimensions.
Simulation results for the two-dimensional
model are summarized in Table \ref{cp2s2}.

\begin{table}
\caption{Simulation: critical parameters for the two-dimensional CP2S. }
\label{cp2s2}
\vspace{.5em}

\begin{tabular}{|c|c|c|c|c|c|c|}
\hline  $\mu$ & $\lambda_{c}$ &$\beta/\nu_\perp$ & $z$ & $\delta$ & $m_p$ & $m_q$
\\ \hline \hline
0.9 & 1.64515(5) & 0.63(5)& 1.95(5) & $> 0.35$& 1.40(2)& 1.52(2)  \\  \hline
0.75 & 1.61640(5) & 0.73(5) & 1.78(6) &0.44(3)  &1.32(3) &  1.33(3) \\  \hline
0.5 & 1.47290(5) & 0.74(3)& 1.72(3) & 0.46(2)& 1.298(8) & 1.322(8)  \\  \hline
0.25& 1.13730(5)& 0.76(2)& 1.73(2) & 0.45(2)& 1.30(2) & 1.31(2)  \\  \hline
0.1 & 0.743160(5)& 0.78(1) & 1.73(2) & 0.46(1)  & 1.305(10)& 1.315(12) \\  \hline \hline
CP/DP & 1.64874(4) & 0.797(3)& 1.7674(6)&0.4523(10) &1.3264(5) &  \\
\hline
\end{tabular}
\end{table}

\section{Field theoretic analysis}

In this section we extend the field theory or continuum representation of DP to the
two-species case, to determine whether the presence of aditional species
changes the scaling behavior.  Since the theory of DP has been known for some time,
we give a bare outline of this analysis, referring the
reader to references \cite{doi,peliti,RDRV,tauber,wijland} for details.
To begin, we modify the lattice model so as to facilitate
the definition of a continuum description following the Doi-Peliti formalism \cite{doi,peliti},
which has been applied to DP in \cite{RDRV} and \cite{wijland}.  (The latter study
applies the Wilson renormalization group to the problem.)

In the Doi-Peliti formalism, the master equation governing the evolution of the probability vector
$| P(t) \rangle \equiv \sum_{\cal C} p({\cal C},t) |{\cal C} \rangle$ (the sum is
over all configurations), is written in the form $d | P \rangle/dt = L |P \rangle$, where
the evolution operator $L$ is
composed of creation and annihilation operators.
Starting from this ``microscopic" description, one derives an effective action ${\cal S}$
via a path-integral mapping.  Then, taking the continuum limit, one arrives at a field
theory for the model.  Of the many lattice models that belong to the DP universality
class, the simplest to analyze in this manner is the Malthus-Verhulst process (MVP).
Here, each site $i$ of a lattice hosts a number $n_i \geq 0$ of particles.  The transitions
at a given site are creation ($n_i \to n_1 + 1$) at rate $\lambda n_i$ and annihilation
($n_i \to n_1 - 1$) at rate $\mu n_i + \nu n_i(n_i-1)$.  In addition, particles hop between
nearest-neighbor sites at rate $D$.

For the MVP on a ring of $\ell$ sites, one has the set of basis
configurations $|n_1,...,n_\ell \rangle$.  Letting $c_i$ and $c_i^\dagger$ denote, respectively,
annihilation and creation operators associated with site $i$, we have, by definition,
$c_i |n_1,...,n_i,...,n_\ell \rangle = n_i |n_1,...,n_i-1,...,n_\ell \rangle$ and
$c_i^\dagger |n_1,...,n_i,...,n_\ell \rangle = |n_1,...,n_i+1,...,n_\ell \rangle$.
Then the evolution operator for the MVP is,

\begin{eqnarray}
\nonumber
L_{MVP}  &=& \large \sum_i \left[ \lambda (c_i^\dagger -1) c_i^\dagger c_i
                   + (1- c_i^\dagger)(\mu + \nu c_i) c_i \right]
\nonumber
\\
                   &+& \frac{D}{2} \sum_i \left[ (c_i^\dagger - c_{i+1}^\dagger) c_{i+1}
                   + (c_i^\dagger - c_{i-1}^\dagger) c_{i-1} \right]
\label{LMVP}
\end{eqnarray}

\noindent Following the steps detailed in \cite{RDRV}, one arrives at the effective action
for the MVP,

\begin{equation}
{\cal S}_{MVP} = \int dt \int dx \left[ \hat{\psi} (\partial_t + w - D \nabla^2) \psi
                 + \nu \hat{\psi} \psi^2 - \lambda \hat{\psi}^2 \psi \right],
\label{smvp}
\end{equation}
\vspace{1em}

\noindent where $w \equiv \mu - \lambda$, the continuum limit has been taken,
and terms higher than third order have
been discarded, as they are irrelevant to critical behavior.
(We recall that $\hat{\psi}(x,t)$ is an auxiliary field that arises in the mapping.  The operator
that governs the evolution of the probability generating function is given by the functional
integral
$U_t = \int {\cal D} \psi \int {\cal D} \hat{\psi} \exp[-{\cal S}(\psi,\hat{\psi})]$;
see \cite{peliti,RDRV}.)

\begin{figure}[h]
\includegraphics[width=10cm,clip=true]{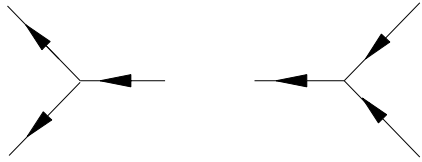}
\vspace{-2.5cm}

\caption{\footnotesize{The two three-field vertices in the field theory of DP.  Lines
exiting a vertex correspond to $\hat{\psi}$ while those entering correspond to $\psi$.}}
\label{vert}
\end{figure}

\begin{figure}[h]
\vspace{2em}
\includegraphics[width=10cm,clip=true]{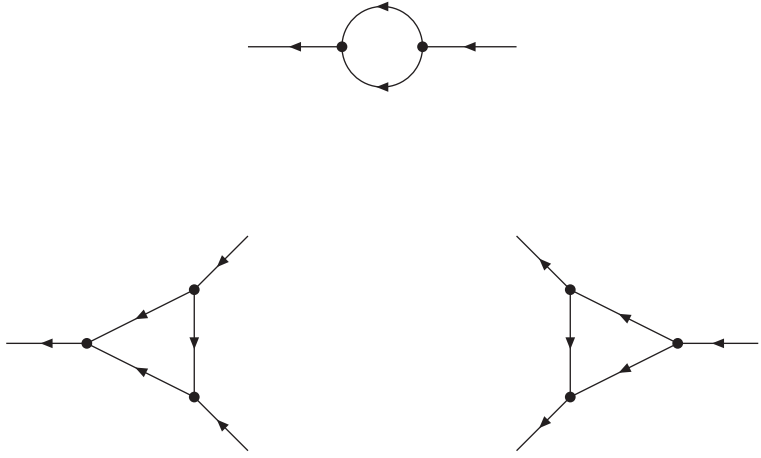}
\caption{\footnotesize{The one-loop diagrams in the field theory of DP, leading to renormalization of
$w$, $\mu$, and $\lambda$, respectively.}}
\label{diags}
\end{figure}

The action of Eq. (\ref{smvp})
is equivalent that of DP, and serves as the starting point for renormalization group (RG)
analyses \cite{janssen,grassberger,wijland}.
(One usually imposes the relation $\nu = \lambda$ via a rescaling
of the fields, but this is not needed here.) In the RG analysis the bilinear term, naturally, defines
the propagator, while the cubic terms correspond to the vertices shown in Fig. \ref{vert}.  These
terms lead, via diagrammatic analysis, to a nontrivial DP fixed point below $d_c = 4$
dimensions.  The one-loop diagrams which yield, to lowest order, the recursion relations
for parameters $w$, $\lambda$, and $\nu$ are shown in Fig. \ref{diags}.

\vspace{2em}

\vspace{2em}

Now consider the two-species CP.  To formulate a minimal field theory, we consider a two-species
MVP; call it MVP2.  Let $m_i$ and $n_i$ denote, respectively, the number of particles of species A and B at
site $i$, and let $a_i$ and $a_i^\dagger$, and $b_i$ and $b_i^\dagger$, denote the associated
annihilation and creation operators.  We require the annihilation rate for species A to be
a decreasing function of $n_i$ and vice-versa; a simple choice for the annihilation rate
of an A particle at site $i$ is $\mu \exp [-\gamma n_i]$, where $\gamma$ is a positive constant,
and similarly for B particles, with $n_i$ replaced by $m_i$.
This corresponds to the evolution operator:

\begin{eqnarray}
\nonumber
L_{MVP2}  &=& \large \sum_i \left[ \lambda (a_i^\dagger -1) a_i^\dagger a_i
                   + (1- a_i^\dagger)(\mu e^{-\gamma b_i\dagger b_i} + \nu a_i) a_i \right]
\nonumber
\\
                   &+& \frac{D}{2} \sum_i \left[ (a_i^\dagger - a_{i+1}^\dagger) a_{i+1}
                   + (a_i^\dagger - a_{i-1}^\dagger) a_{i-1} \right]
\nonumber
\\
          &+& \large \sum_i \left[ \lambda (b_i^\dagger -1) b_i^\dagger b_i
                   + (1- b_i^\dagger)(\mu e^{-\gamma a_i\dagger a_i} + \nu b_i) b_i \right]
\nonumber
\\
                   &+& \frac{D}{2} \sum_i \left[ (b_i^\dagger - b_{i+1}^\dagger) b_{i+1}
                   + (b_i^\dagger - b_{i-1}^\dagger) b_{i-1} \right]
\label{LMVP2}
\end{eqnarray}

\noindent To avoid ambiguity, we interpret the exponentials as being in
{\it normal order}, i.e., all creation operators to the left of annihilation operators.
Recalling that terms with four or more fields are irrelevant, we may
expand the exponentials, retaining only the terms $\propto b_i^\dagger b_i$ and
$\propto a_i^\dagger a_i$.  Using $:X:$ to denote the normal-ordered expression of
$X$, it is straightforward to show that

\begin{equation}
:e^{-\gamma b^\dagger b}: = 1 - (1-e^{-\gamma}) b^\dagger b +  I \equiv 1 - \bar{\gamma} b^\dagger b +  I,
\label{normo}
\end{equation}

\noindent where $I$ consists of terms with four or more operators.  (With the truncation comes the
possibility of a negative rate, but this is of no consequence in the RG analysis.)
Now, following the usual procedure, we obtain the effective action for the two-species MVP:

\begin{eqnarray}
{\cal S}_{MVP2} &=& \int dt \int dx \left[ \hat{\psi} (\partial_t + w - D \nabla^2) \psi)
                 + \nu \hat{\psi} \psi^2 - \lambda \hat{\psi}^2 \psi \right]
\nonumber
\\
                &+& \int dt \int dx \left[ \hat{\varphi} (\partial_t + w - D \nabla^2) \varphi)
                 + \nu \hat{\varphi} \varphi^2 - \lambda \hat{\varphi}^2 \varphi \right]
\nonumber
\\
                &-& \bar{\nu} \int dt \int dx \left[ \hat{\varphi} \varphi \psi
                 + \hat{\psi} \psi \varphi \right],
\label{smvp2}
\end{eqnarray}

\noindent where $\bar{\nu} = \bar{\gamma} \mu$.  Here $\psi$ and $\hat{\psi}$ are fields
associated with species A; $\varphi$ and $\hat{\varphi}$ are associated with species B.
The first two lines of the above expression correspond to independent MVPs;  the third
represents the symbiotic interaction between them.
[While such a minimal action could have been ``postulated" directly, we prefer to
start with the microscopic expression of Eq. (\ref{LMVP2}), since it describes a valid stochastic process.]

There are two cubic terms in the action involving only species A (i.e., the vertices shown in Fig. \ref{vert}),
two involving only B (those of Fig. \ref{vert} drawn, say, with broken lines) and two
vertices with a {\it mixed pair} of incoming lines, and a single outgoing line, which may belong to
either species.  One readily identifies the one-loop diagrams leading to renormalization of
the parameter $\bar{\nu}$.  On the other hand, {\it no}
diagrams (at any order) involving mixed-species vertices can affect the recursion
relations for the DP parameters
$w$, $\nu$, and $\lambda$.  The reason is that the presence of a mixed-species vertex
anywhere in a diagram implies that
the lines entering the diagram are mixed, so that it can only contribute to
the recursion relation for $\bar{\nu}$.  We conclude that the interaction between species cannot
alter the scaling behavior, which must therefore remain that of DP.  At one-loop order, there are
two fixed-point values for $\bar{\nu}$, namely, $2 \lambda$ and zero, the latter corresponding to
independent processes.

\section{Conclusions}

We study symbiotic interactions in contact-process-like models in one and two dimensions.
For this purpose, we propose a
two-species model (CP2S), in which the death rate is reduced (from unity to
$\mu$), on sites occupied by both species.
A related model (CPSLS), in which each species is confined to its own
sublattice, is also studied in one dimension, and found to exhibit similar behavior.
Simulations reveal that the phase transition between active and absorbing states
is continuous, and that the critical creation rate $\lambda_c$ is reduced in the presence
of symbiosis.  This means that the loss of one species will rapidly lead to extinction, since
the system is then a basic contact process operating at $\lambda < \lambda_c$.
Although this might suggest identifying the density $q$ of doubly occupied sites as the
order parameter,  we find that the particle density $\rho$ (which includes a large
contribution from singly occupied sites), scales in the same manner as $q$.

Mean-field theory (in both the one- and two-site approximations),
predicts a discontinuous phase transition in any number of
dimensions, for $\mu$ sufficiently small.
A discontinuous transition between an active and an absorbing phase is not
expected in one-dimensional systems of the kind studied here \cite{hinrichsen1d},
nor do our simulations show any evidence of a discontinuous transition in two dimensions.
Nevertheless, we cannot discard the possibility of such a transition for
$d\geq 2$, for small values of $\mu$,
or under rapid particle diffusion, which generally favors mean-field-like behavior.

Overall, the critical behavior of the symbiotic models is consistent with that of directed
percolation.  Corrections to scaling are, however, more significant than in the basic CP,
so that a study restricted to smaller systems, or to only one
kind of simulation, could easily suggest non-DP behavior.
These corrections are stronger, and of longer duration, the {\it smaller} the intensity
of symbiosis.  Thus, in the two-dimensional case, the decay of $\rho$ (in initial-decay
studies) attains the expected power-law regime (with a DP value for the decay exponent),
{\it except} for $\mu = 0.9$, the weakest symbiosis studied.
A similar tendency is observed in the QS simulations of the one-dimensional CP2S, for
which the estimates for critical exponents and the critical moment ratio
$m_c$ differ most from DP values for $\mu = 0.9$.

In the initial-decay studies in one dimension, for smaller intensities of symbiosis
(i.e., $\mu = 0.9$ and 0.75), we observe anomalous growth
of fluctuations in the order parameter.  The latter are characterized
by $m_\rho - 1 = \mbox{var}(\rho)/\rho^2$,
which is expected to grow $\sim t^{1/z}$, before saturating at its QS value.
The growth at long times corresponds to a $z$ value significantly smaller than that of DP.
The exponent $z_q$ associated with the growth of $m_q$ is substantially larger, though
still slightly below the DP value.  In contrast with these anomalies, the spreading
exponents are found to take DP values in one dimension, {\it independent} of the
degree of symbiosis.  Thus we are inclined to regard the asymptotic scaling of the
symbiotic models as being that of DP, and to interpret the deviations as arising from
finite-time and finite-size corrections.
One might conjecture that under strong symbiosis, the critical system is rapidly
attracted to the DP fixed point (although not as rapidly as is the basic CP),
whereas for weak symbiosis, it makes a long
excursion into a regime in which DP-like scaling is not evident, before finally
returning to the vicinity of the DP fixed point.  The asymptotic scaling behavior is presumably
associated with large, sparsely populated but highly correlated regions of doubly
occupied sites, which, for reasons of symmetry, behave analogously to DP space-time clusters.
The presence of isolated particles, which are relatively numerous and long-lived
for weak symbiosis, could mask the asymptotic critical behavior, on short scales.
We defer further analysis of these questions to
future work.

Extending the field theory of DP to the two-species case, we find that the irrelevance of
four-field terms
makes DP extremely robust, since the only possible three-field vertices are already present
in the single-species theory. This means
that the interaction between species cannot alter the scaling behavior, as already noted
by Janssen in the case of multi-species DP processes \cite{janssen01}.
Our simulation results, as noted, support this conclusion.    A more detailed field-theoretic analysis,
including the evolution of the lowest order irrelevant terms, might shed some light on the
scaling anomalies observed in the simulations.
\vspace{1cm}

\noindent{\bf Acknowledgments}
\vspace{1em}

We are grateful to Miguel A. Mu\~noz for helpful comments.
This work was supported by CNPq and FAPEMIG, Brazil.

\bibliographystyle{apsrev}

\end{document}